\documentclass[acmsmall]{acmart}
\usepackage{listings}
\usepackage{subfigure}

\newcommand{\scientific}[2]{\ensuremath{{#1}\times 10^{#2}}}

\title{Replicated Computational Results (RCR) Report for ``Code Generation for Generally Mapped Finite Elements''}
\keywords{Finite Element, Partial Differential Equations, Replicated Computational Results}

\author{Neil Lindquist}
\orcid{0000-0001-9404-3121}
\affiliation{%
    \position{Research Assistant} 
    \institution{Saint John's University}
    \city{Collegeville}
    \state{MN}
    \country{USA}
}
\email{nslindquist@csbsju.edu}

\begin{document}

\setcopyright{acmcopyright}
\acmJournal{TOMS}
\acmYear{2019} \acmVolume{1} \acmNumber{1} \acmArticle{1} \acmMonth{1} \acmPrice{15.00}

\begin{CCSXML}
<ccs2012>
<concept>
<concept_id>10002944.10011123.10011676</concept_id>
<concept_desc>General and reference~Verification</concept_desc>
<concept_significance>500</concept_significance>
</concept>
<concept>
<concept_id>10002950.10003705.10011686</concept_id>
<concept_desc>Mathematics of computing~Mathematical software performance</concept_desc>
<concept_significance>300</concept_significance>
</concept>
<concept>
<concept_id>10002950.10003705.10003707</concept_id>
<concept_desc>Mathematics of computing~Solvers</concept_desc>
<concept_significance>100</concept_significance>
</concept>
</ccs2012>
\end{CCSXML}

\ccsdesc[500]{General and reference~Verification}
\ccsdesc[300]{Mathematics of computing~Mathematical software performance}
\ccsdesc[100]{Mathematics of computing~Solvers}

\begin{abstract}
    ``Code Generation for Generally Mapped Finite Elements'' includes performance results for the finite element methods discussed in that manuscript.
    The authors provided a Zenodo archive with the Firedrake components and dependencies used, as well as the scripts that generated the results.
    The software was installed on two similar platforms; then, new results were gathered and compared to the original results.
    After completing this process, the results have been deemed replicable by the reviewer.
\end{abstract}
	
\maketitle

\section{Introduction}
The results replication effort for ``Code Generation for Generally Mapped Finite Elements'' by Robert Kirby and Lawrence Mitchell focused on Section~4 of the manuscript, which provides example results and comparisons between the newly implemented finite element transformations and the existing Lagrange transformation.
The original results were obtained using an Intel E5-2640v3 processor; however, the replication was done using two different machines, an Intel E5-2698v4 and an Intel E5-2609v4, designated \texttt{node0} and \texttt{node1}, respectively.

\section{Replicating the Results}
The exact version of the software used to generate the results was archived on Zenodo~\cite{zenodo}.
This archive contains the versions of almost all of the Firedrake components used, the scripts that generate the data, and the resulting data files.
The one dependency not included in the archive is SciPy, which is required to generate the sparsity graphs.
The original archive had a few configuration issues, but the authors worked with the reviewer to correct the issues, which resulted in the aforementioned version of the archive.

\subsection{Installation}
Because Firedrake is able to configure an installation based on the Zenodo archive, installation was straight forward.
The machines used to replicate the results shared a file system and operating system image,  and so, Firedrake was installed from \texttt{node0} then reused from \texttt{node1}.
The installation script for Firedrake normally ensures the correct system dependencies are present using \texttt{apt-get}.
However, \texttt{apt-get} was not present on the replication machines.
So, the system dependencies had to be manually verified, including
\begin{itemize}
    \item GNU make 3.82
    \item cmake 2.8.12.2
    \item gcc and gfortran 4.8.5
    \item MPICH 3.2
    \item Python 3.7.3
\end{itemize}

Installation was then accomplished by running the following commands from a bash shell.
Note that the \texttt{--no-package-manager} and \texttt{--disable-ssh} flags were added by the reviewer to the recommended installation.
The \texttt{--no-package-manager} was required since \texttt{apt-get} was not present on the replication machines.
Git was not configured correctly to clone repositories over ssh.
So, the \texttt{--disable-ssh} flag was added to use https from the start, instead of attempting to use ssh first for every repository.
\begin{lstlisting}[breaklines,
                    language=bash,
                    showstringspaces=false]
$ export PETSC_CONFIGURE_OPTIONS="--download-pastix --download-ptscotch"
$ python3 firedrake-install --doi 10.5281/zenodo.3234784 --slepc --no-package-manager --disable-ssh --install git+https://github.com/thomasgibson/scpc.git@3a1173ebb3610dcdf5090088294a542274bbb73a 
$ chmod +x firedrake/bin/activate
$ firedrake/bin/activate
$ firedrake/bin/pip install scipy
\end{lstlisting}

\subsection{Execution}

The various results were generated from the 13 python scripts in the Zenodo archive.
On each machine, first \verb|firedrake/bin/activate| was run in the bash shell to ensure the correct virtual environment was used.
Then, each of the scripts was run using the Firedrake installation's python.
Note that both \verb|chladni.py| and \verb|guitar.py| both have command line configurable options.
The default values matched those described in the paper, except for the number of eigenvalues to generate.
So, \verb|-eps_nev 30| was passed to those scripts.

The output was a mixture of pdf and csv files.
The correspondence between the output files and the results presented in the manuscript is obvious except for the timing results.
For the timing results, each discretization has its own csv file containing the solution error and various execution times for each mesh size.
The timing results presented in the manuscript are the time to assemble the discretization, in column \texttt{SNESJacobianEval}, the time to factor and solve the linear system, in column \texttt{KSPSolve}, and the total solver time, in column \texttt{SNESSolve}.

\section{Evaluation Of Replicated Results}
There were two problems used to compare the newly implemented transformations with the existing transformation, the second degree Poisson's equation and the fourth degree biharmonic equation.
Because different hardware was used for the original results and the replicated results, the timing results were only judged on whether the relative performances are similar. 
For the other results, only differences that can be explained by round off error was considered acceptable. 
Additionally, there were two example problems, the Cahn-Hilliard equation and the Chladni plate problem.
These example problems each generate a set of images.
These images were expected to look similar, with minor differences considered acceptable.

\subsection{Comparison of Model Problems}
%TODO this paragraph feels small
  % add transition off the end?
The FLOP counts and sparsity for assembling the linear systems should both be deterministic computations with limited floating point operations, so both should be highly replicable.
% Flops chart
The FLOP counts generated by each machine are exactly equal to those provided in the archived data and match up with the graphs in the manuscript.
% Sparsity graphs
Similarly, the generated sparsity patterns were visually indistinguishable with the ones in the archive and the manuscript.

% Error
Figure~9a in the manuscript contains the \(L^2\) error of the Poisson equation.
The \(L^2\) errors in the generated results were all within \scientific{1.75}{-14} of the archived data.
Similarly, for the \(L^2\) error of the biharmonic equation, presented in Figure~9b in the manuscript, the errors of the generated results were all within \scientific{3.25}{-15} of the archived data, except for the error of \(P^5\) when \(N=2^7\) which was off by \scientific{6.87}{-13}.
These errors are reasonable for accumulated floating point round off error, and so are considered replicated.

\begin{figure}
    \includegraphics[width=0.35\textwidth]{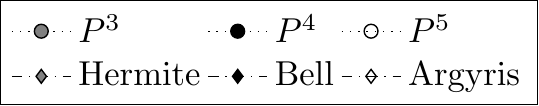}

    \subfigure[][Assembly time of the Laplace system on \texttt{node0}.]{%
        \includegraphics[width=0.45\textwidth]{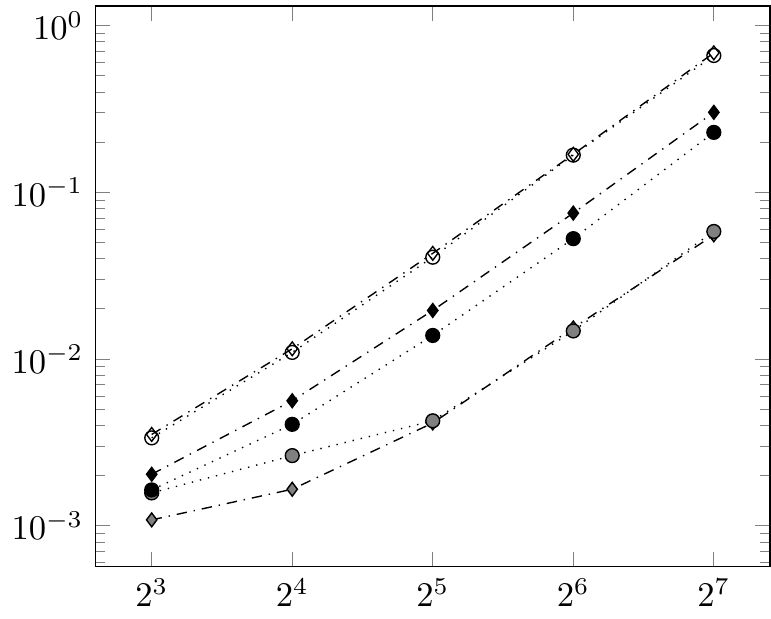}
    }
    \subfigure[][Assembly time of the Laplace system on \texttt{node1}.]{%
        \includegraphics[width=0.45\textwidth]{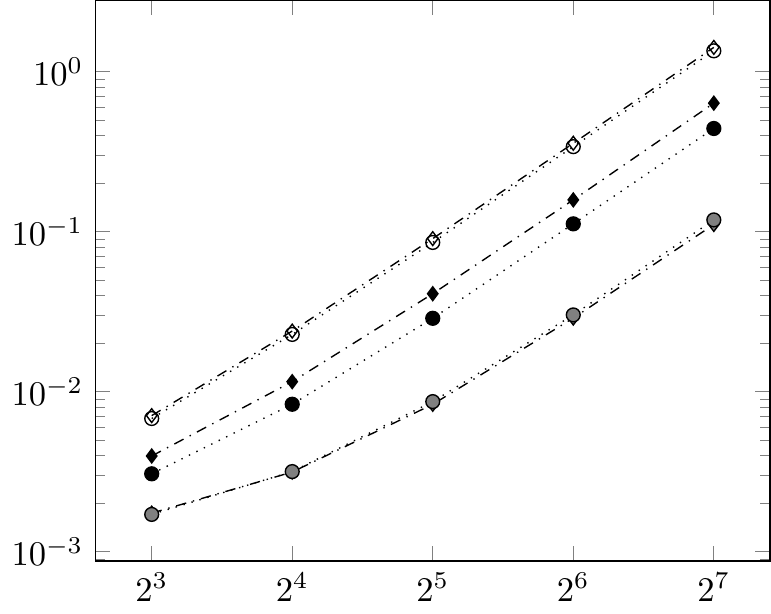}
    }
    \caption{Replication of Laplace assembly time from Figure 11a in the manuscript.}
    \label{fig:assembly-poisson}
\end{figure}

\begin{figure}
    \includegraphics[width=0.45\textwidth]{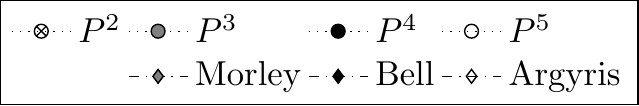}
        
    \subfigure[][Assembly time of the biharmonic system  on \texttt{node0}.]{%
        \includegraphics[width=0.45\textwidth]{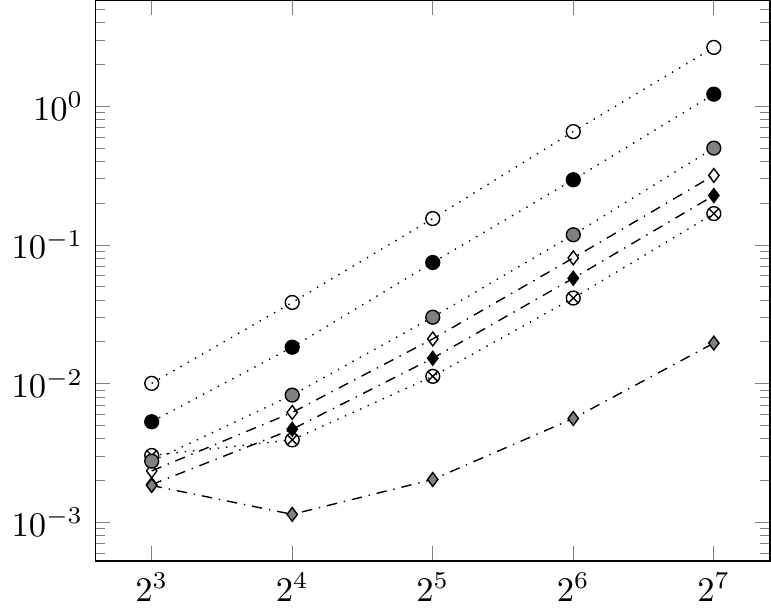}
    }
    \subfigure[][Assembly time of the biharmonic system  on \texttt{node1}.]{%
        \includegraphics[width=0.45\textwidth]{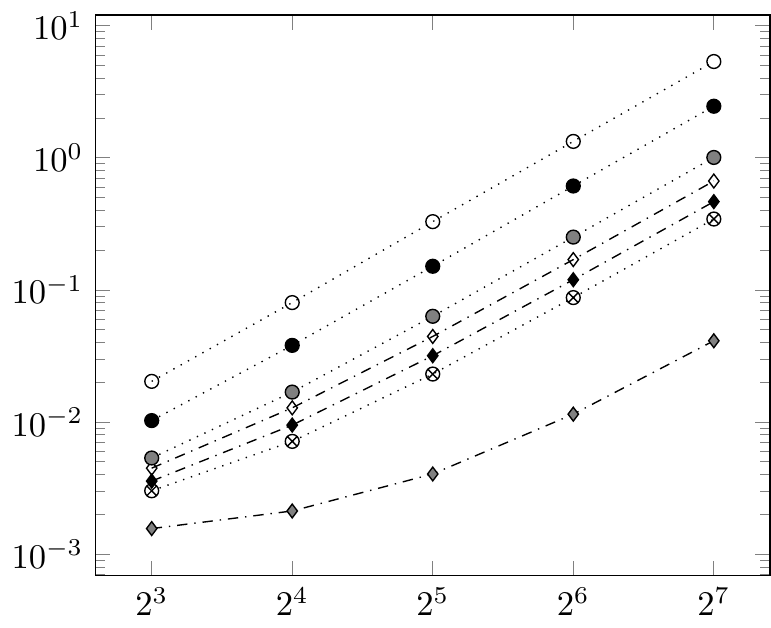}
    }
    \caption{Replication of biharmonic assembly time from Figure 11b in the manuscript.}
    \label{fig:assembly-biharmonic}
\end{figure}

% Assembly time
% SNESJacobianEval
Figure~11 in the manuscript shows the time to assemble the Laplace and biharmonic operators.
The manuscript notes that for the Poisson system ``Hermite and \(P^3\), and Argyris and \(P^5\) elements require very similar assembly time, with Bell somewhat higher than \(P^4\)''.
Figure~\ref{fig:assembly-poisson} shows the replication performance of assembling the Laplace operator.
For the biharmonic operator, it is noted that a ``clear win for \(H^2\) elements'' is shown, with the replication performance shown in Figure~\ref{fig:assembly-biharmonic}.
Both sets of replication results agree with the manuscripts claims of relative performance.
Additionally, the graphs look similar with the only difference is that the time to assembly of the \(H^2\) elements is slightly lower, which is acceptable.

\begin{figure}
    \includegraphics[width=0.35\textwidth]{poisson-basic-legend.pdf}
    
    \subfigure[][The time to factor and solve the linear system on \texttt{node0}.]{%
        \includegraphics[width=0.45\textwidth]{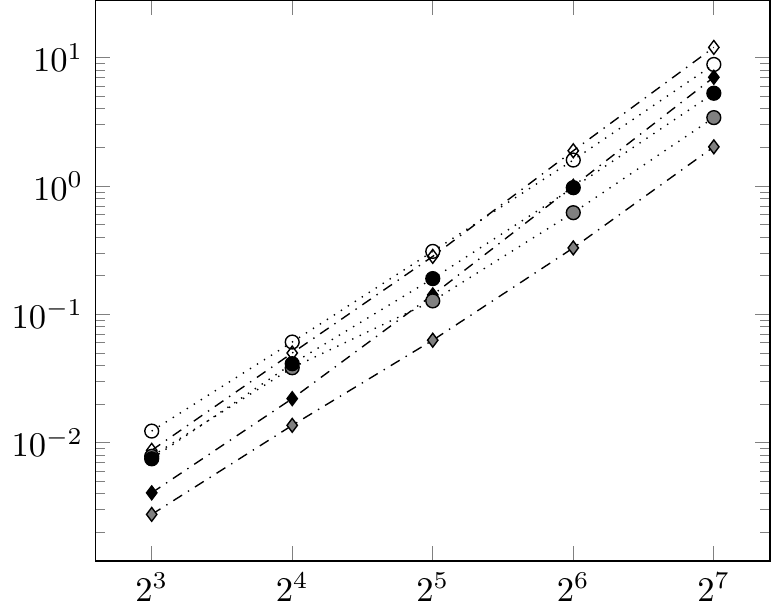}
    }
    \subfigure[][The time to factor and solve the linear system on \texttt{node1}.]{%
        \includegraphics[width=0.45\textwidth]{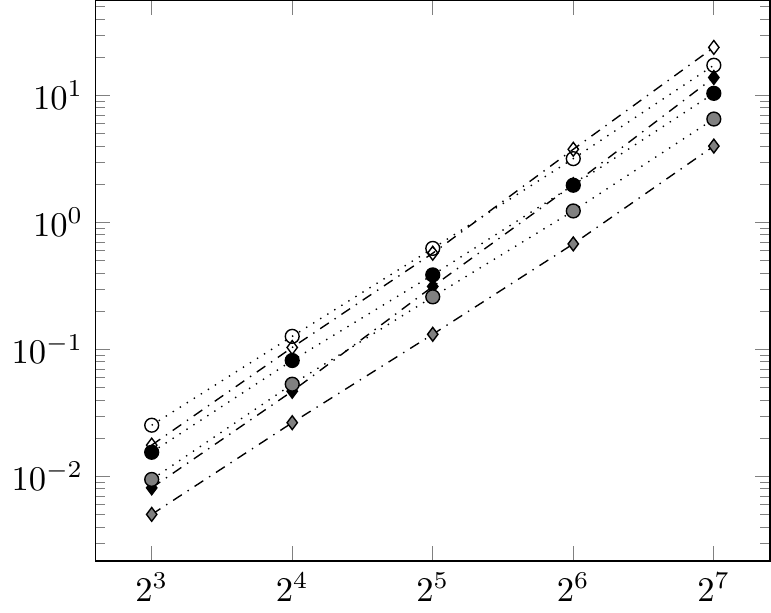}
    }
    \caption{Replication of the time to factor and solve the sparse linear system for the Laplacian operator, as per Figure~14a.}
    \label{fig:factorsolve-poisson}
\end{figure}

\begin{figure}
    \includegraphics[width=0.6\textwidth]{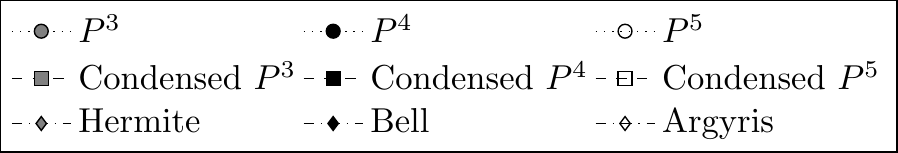}
    
    \subfigure[][The total solver time on \texttt{node0}.]{%
        \includegraphics[width=0.45\textwidth]{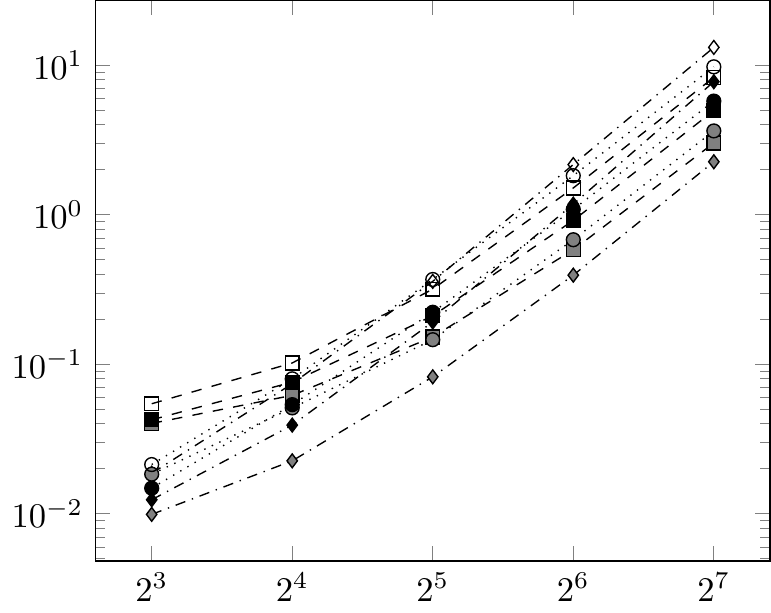}
    }
    \subfigure[][The total solver time on \texttt{node1}.]{%
        \includegraphics[width=0.45\textwidth]{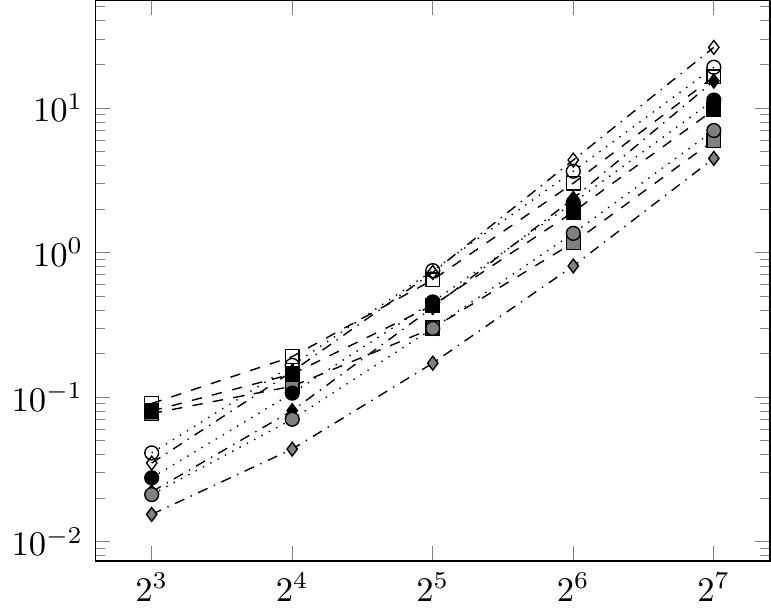}
    }
    \caption{Replication of the total solver time for the Poisson equation, as per Figure~14b of the manuscript.}
    \label{fig:total-poisson}
\end{figure}

% Factor and solve time
% KSPSolve
%
% Total time
% SNESSolve
In the manuscript, Figure~14 shows the time to factor and solve the linear system, and the overall solver time for the Poisson equation.
For the Poisson equation, the manuscript notes that ``Hermite and Bell systems are actually cheaper to solve than any of the \(P^k\) systems, and Argyris appears to have a comparable cost to \(P^4\) rather than \(P^5\)'' and that ``condensation improves the total run-time for \(P^k\) elements on the finer meshes''.
For the Poisson problem, the time to factor and solve the linear system is shown in Figure~\ref{fig:factorsolve-poisson}, and the overall solver time is shown in Figure~\ref{fig:total-poisson}.
As can be seen in the figures, the claims about the performance of the Argyris and Bell systems for the Poisson equation do not hold in the replicated data.
However, the rest of the performance claims do hold, and the graphs otherwise look similar.
Note that the time to factor and solve the linear equation makes up almost the entirety of the total solve time.
Thus, this difference in performance likely comes from differences in either the linear systems or the configuration and runtime of the linear solver.
Because the previous parts of the comparison were replicated successfully, a difference in the runtime of the linear solver is more likely the cause of this issue.
Furthermore, the conclusion only states that the ``numerical results indicate that these elements are viable'' and ``of comparable cost to deploy as any Lagrange elements,'' meaning that the replicated performance still supports those conclusions.
Hence, the performance results were considered to be successfully replicated.

\begin{figure}
    \includegraphics[width=0.45\textwidth]{biharmonic-legend.pdf}
    
    \subfigure[][The time to factor and solve the linear system on \texttt{node0}.]{%
        \includegraphics[width=0.45\textwidth]{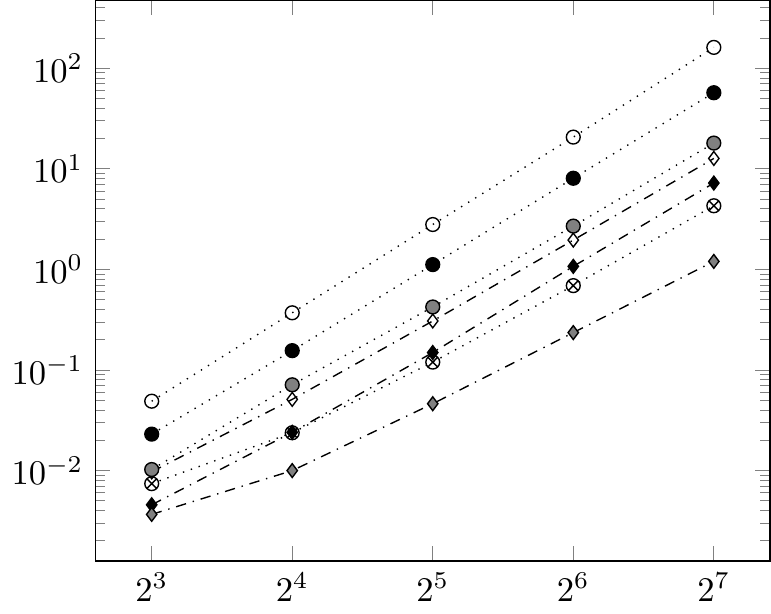}
    }
    \subfigure[][The time to factor and solve the linear system on \texttt{node1}.]{%
        \includegraphics[width=0.45\textwidth]{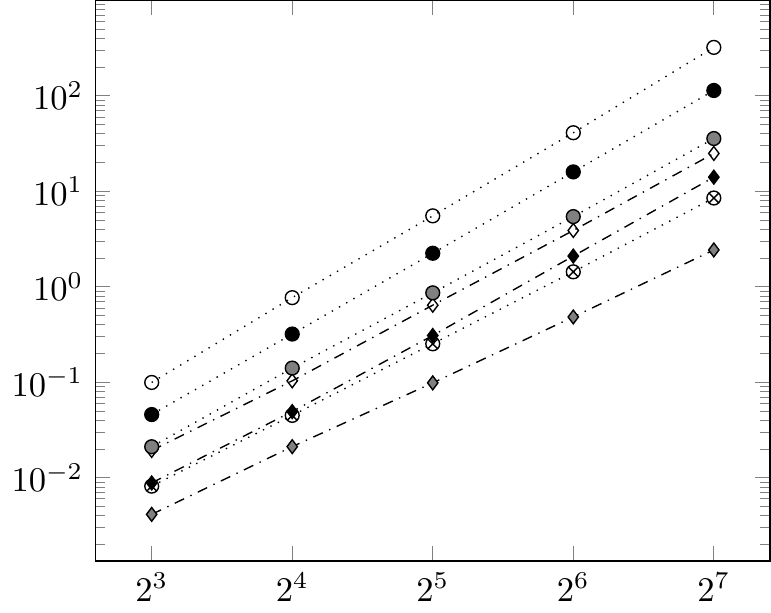}
    }
    \caption{Replication of the time to factor and solve the sparse linear system for the biharmonic operator, as per Figure~15a, respectively.}
    \label{fig:factorsolve-biharmonic}
\end{figure}

\begin{figure}
    \includegraphics[width=0.45\textwidth]{biharmonic-legend.pdf}
    
    \subfigure[][The total solver time on \texttt{node0}.]{%
        \includegraphics[width=0.45\textwidth]{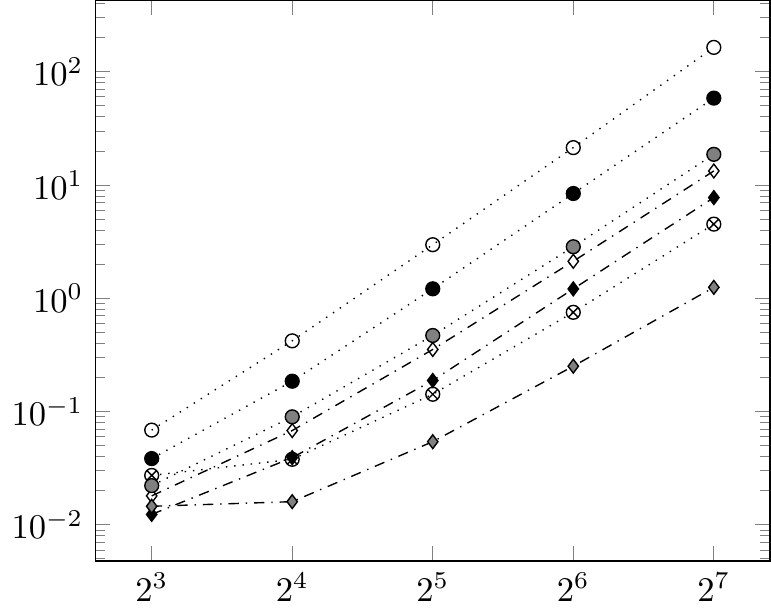}
    }
    \subfigure[][The total solver time on \texttt{node1}.]{%
        \includegraphics[width=0.45\textwidth]{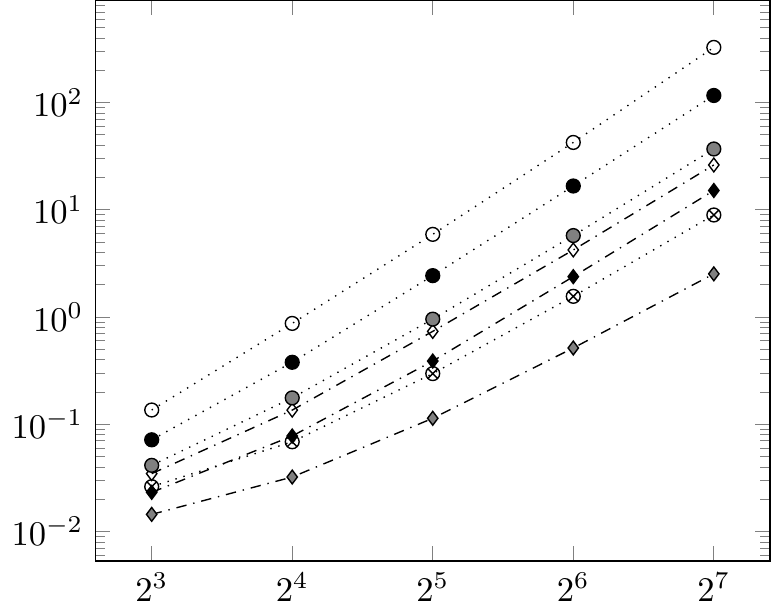}
    }
    \caption{Replication of the total solver time for the biharmonic equation, as per Figure~15b of the manuscript.}
    \label{fig:total-biharmonic}
\end{figure}

Similarly, Figure~15 in the manuscript shows the time to factor and solve the linear system, as well as the overall solver time, for the biharmonic equation.
The manuscript merely states that ``solution of the Argyris element typically being cheaper than even the \(P^3\) interior penalty method, not to mention \(P^5\).''
The time to factor and solve the linear system is shown in Figure~\ref{fig:factorsolve-biharmonic}, and the overall solver time is shown in Figure~\ref{fig:total-biharmonic}
Unlike the performance of the Laplacian operator, the biharmonic operator performed close to the reported results.

\subsection{Example Problems}
The Cahn-Hillard example produced only the images of the initial and final states.
Additionally, the initial state is generated with a fixed randomization seed, so directly comparing the resulting images is possible. 
The initial state and final state images generated by each machine are indistinguishable from the respective images in the Zenodo archive and in the manuscript.

\begin{figure}
    \subfigure[][The image in the Zenodo archive and Figure~18 of the manuscript.]{%
        \includegraphics[width=0.3\textwidth]{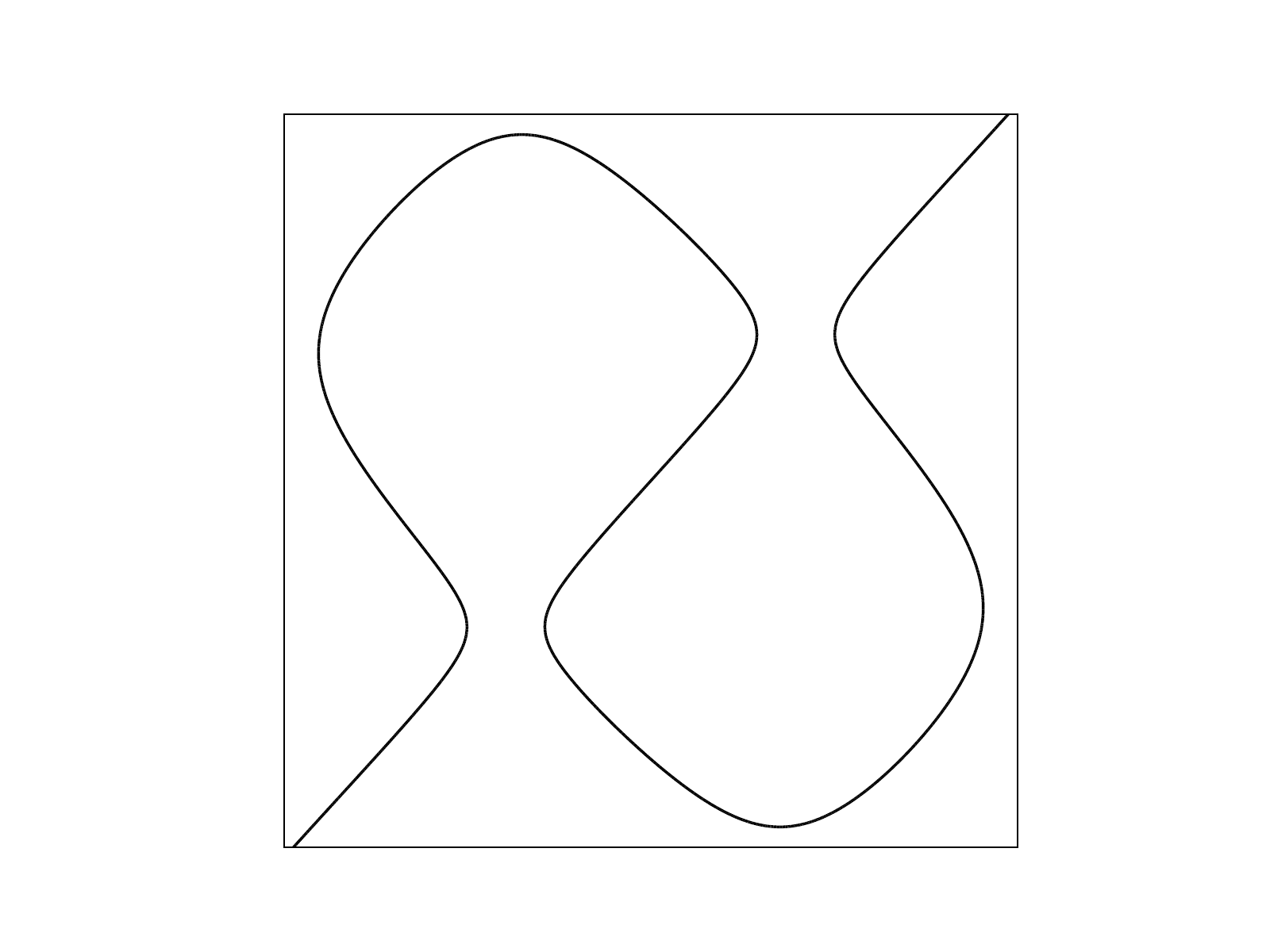}
    }
    \subfigure[][The image generated by \texttt{node0}.]{%
        \label{fig:Chladni-eigfunc8-node0}
        \includegraphics[width=0.3\textwidth]{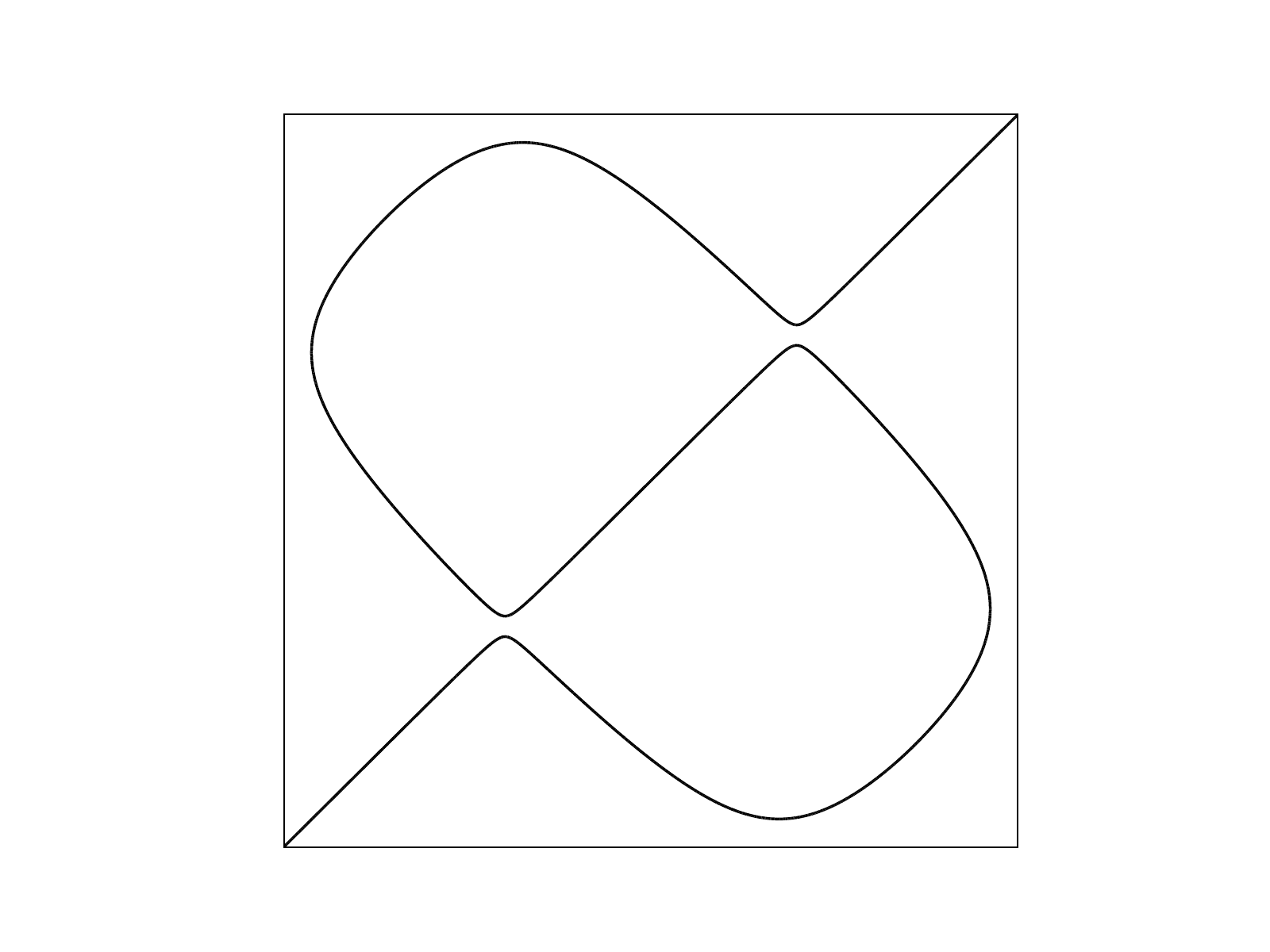}
    }
    \subfigure[][The image generated by \texttt{node1}.]{%
        \label{fig:Chladni-eigfunc8-node1}
        \includegraphics[width=0.3\textwidth]{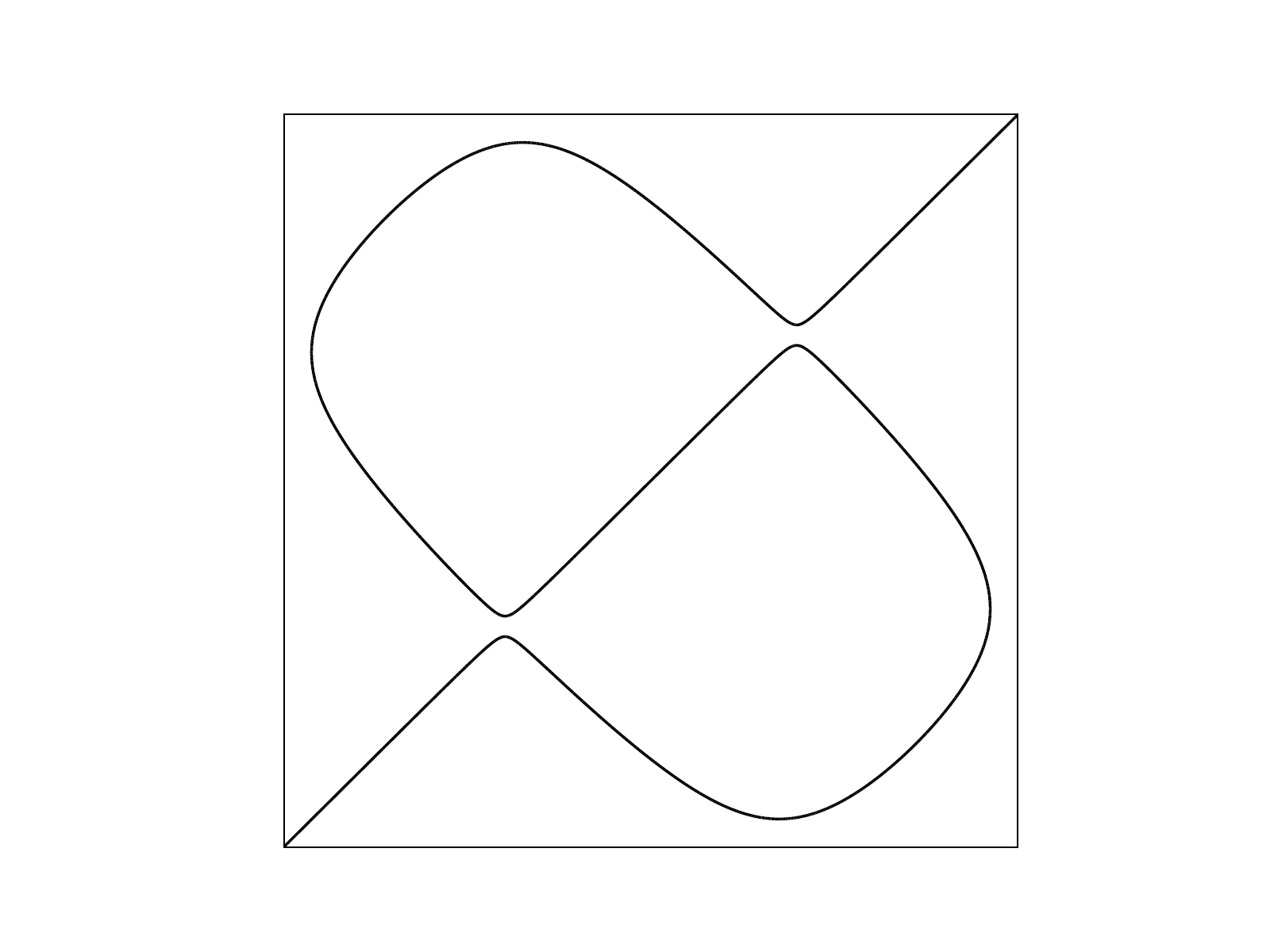}
    }
    \caption{The various images for the discussed eigenfunction.}
    \label{fig:Chladni-eigfunc8}
\end{figure}

The Chladni plate problems each provide a series of images that depict various Chladni figures for the respective plate shapes.
For the square plate, both machines generated almost the same set of eigenpairs with a few minor differences from the versions in the archive and manuscript.
First, there are three eigenfunctions with eigenvalues of 0.
These eigenfunctions are present in the archived results but were not generated by either replication machine.
However, these eigenfunctions are not show in the manuscript, so this issue was not considered to affect the replicability.
Second, two of the eigenfunctions rendered slightly differently than are shown in the manuscript and archived data.
These eigenfunctions correspond to the same eigenvalue and are rotations of each other, so only one is discussed here. 
Figure~\ref{fig:Chladni-eigfunc8} shows both of the generated images and the archived image for one of those eigenfunctions.
Note that the exact value of this eigenfunction is similar to the images presented in Figures~\ref{fig:Chladni-eigfunc8-node0} and~\ref{fig:Chladni-eigfunc8-node1}, except the diagonal is continuous~\cite{Ritz:1909:square-plate}.
Given that the replicated result is more accurate that the image in the manuscript, this difference was not considered to affect the replicability of the manuscript.

In addition to computing the Chladni figures on a square plate, figures were computed for a guitar shaped region.
The Zenodo archive only contains images for the first 28 eigenfunctions, which are shown in the manuscript.
However, with the \verb|-eps_nev 30| flag, there are an extra 6 eigenfunctions computed that were not present in the archive; these images match between the two replication machines but were otherwise ignored for this review.
For the other 28 eigenfunctions, both sets of generated images match the images in the archive and manuscript.

\section{Concluding Remarks}
The reviewer attempted to replicate the figures and results in Section~4 of the manuscript.
The installation and execution of the software was straightforward using the Zenodo archive listed in the manuscript.
Furthermore, the results generated were reasonably close to those in the manuscript.
Thus, the reviewer deemed the published results to be replicable.

\begin{acks}
The reviewer would like to thank Lawrence Mitchell for providing guidance on installing and using the software, and Michael Heroux for providing guidance on the RCR review process.
\end{acks}

\bibliography{bibliography}

\end{document}